\documentclass[12pt, preprint]{aastex}
\usepackage{graphicx,xfrac,amsmath,longtable,color}

\def\la{\mathrel{\hbox{\rlap{\hbox{\lower4pt\hbox{$\sim$}}}\hbox{$<$}}}}
\def\ga{\mathrel{\hbox{\rlap{\hbox{\lower4pt\hbox{$\sim$}}}\hbox{$>$}}}}

\def\kms{km~s$^{-1}$}
\def\cms{$\rm{cm\,s^{-1}}$}
\def\cm2g{cm$^{2}$~g$^{-1}$}

\def\magd{mag~d$^{-1}$}

\def\dm15{{$\Delta$}$m_{15}$}

\def\v10{$V_{10}$(Si~II)}

\def\W575{$W(5750)$}
\def\W610{$W(6100)$}
\def\6100{the 6100~\AA\ absorption}

\def\msun{~M$_\odot$}

\def\ni{$^{56}$Ni}
\def\co{$^{56}$Co}

\def\CaII7291{Ca {\sc II}] $\lambda\lambda$7291,7323\ }

\def\OI6300{[O {\sc I}] $\lambda\lambda$ 6300,6364\ }

\def\OIB{[O~{\sc I}] $\lambda$7774}

\newcommand\gr{$\gamma$-ray}
\newcommand\grs{$\gamma$-rays}

\def\apj{Ap. J.}
\def\apjl{Ap. J. Lett.}
\def\apjs{Ap. J. Supp.}
\def\aj{Astron. J.}

\def\mnras{Mon. Not. Royal. Ast. Soc.}
\def\aap{Astron. \& Astroph.}
\def\pasp{Pub. Astr. Soc. Pac.}

\begin{document}

\title{Analysis of Late--time Light Curves of Type IIb, Ib and Ic Supernovae}

\author{J. Craig Wheeler\altaffilmark{1}, V. Johnson\altaffilmark{1}, A. Clocchiatti\altaffilmark{2}}
\authoremail{wheel@astro.as.utexas.edu}
\altaffiltext{1}{Department of Astronomy, University of Texas at Austin,
Austin, TX, USA.}
\altaffiltext{2}{Universidad Catolica, Chile}

\begin{abstract}

The shape of the light curve peak of radioactive--powered core--collapse ``stripped--envelope" supernovae 
constrains the ejecta mass, nickel mass, and kinetic energy by the brightness and diffusion time for 
a given opacity and observed expansion velocity. Late--time light curves give constraints on the ejecta 
mass and energy, given the gamma--ray opacity. Previous work has shown that the principal light curve peaks 
for SN~IIb with small amounts of hydrogen and for hydrogen/helium--deficient SN~Ib/c are often rather 
similar near maximum light, suggesting similar ejecta masses and kinetic energies, but that late--time
light curves show a wide dispersion, suggesting a dispersion in ejecta masses and kinetic energies.
It was also shown that SN~IIb and SN~Ib/c can have very similar late--time light curves, but different 
ejecta velocities demanding significantly different ejecta masses and kinetic energies. We revisit 
these topics by collecting and analyzing well--sampled single--band and quasi--bolometric light curves 
from the literature. We find that the late--time light curves of stripped--envelope core--collapse supernovae 
are heterogeneous. We also show that the observed properties, the photospheric velocity at peak, the 
rise time, and the late decay time, can be used to determine the mean opacity appropriate to the peak. 
The opacity determined in this way is considerably smaller than common estimates. We discuss how 
the small effective opacity may result from recombination and asymmetries in the ejecta. 


\end{abstract}



\section{Introduction}

Clocchiatti \& Wheeler (1997) noted several apparent properties of ``stripped--envelope" supernovae: 
(1) Some events completely trap gamma--rays for hundreds of days. The only known examples of these slow 
light curves were SN~Ib. (2) Some events display a similar photometric evolution with an intermediate later time 
rate of decline, similar to SN~1993J, despite their spectral variety, SN~IIb, SN~Ib, SN~Ic. (3) Some events decline 
especially rapidly from maximum, show an especially large peak to tail contrast, but show a slope comparable to 
those of the intermediate slope class (point 2) at very late times, over 100 d after maximum. These events
all seemed to be SN~Ic. Drout et al. (2011) found that the behavior of the light curve peaks of stripped--envelope 
events were rather similar, with SN~Ib and SN~Ic having statistically indistinguishable decline rates up to 
$\sim$ 40 d after maximum, implying similar ejecta mass and energy. Taddia et al. (2014) find that SN~Ic 
and SN~Ic--BL have shorter rise times than SN~IIb and SN~Ib, but that the immediate post--peak declines 
are similar for all categories. Similar decline rates shortly after peak, with some dispersion, were also noted 
by Cano (2013) and Lyman et al. (2014). This evidence for similar post--peak decline rates belies the 
long--standing evidence for late--time light curve dispersion. Here, we compile light curve data for late--time 
light curves of stripped--envelope supernovae and re-investigate the uniformity of the late--time behavior. 
We concentrate here on light curve properties. See Modjaz et al. (2014) for a recent compilation of spectral 
properties and \citet{Bianco14} for a compilation of light curves. Section 2 defines the analytical basis of our analysis. 
Section 3 presents the data. Section 4 gives a discussion and Section 5 summarizes our conclusions.

\section{Peaks and Tails}

We analyze data from the literature on the light curves of stripped--envelope supernovae. We did our own
compilation, but see Cano (2013) for an independent compilation, including some long--term data of
the same events we present here. We estimate the ejecta mass and energy from the observed rise time 
and photospheric velocity near peak light, 
and an effective opacity, assumed to be constant. Following Arnett (1982) we write
\begin{equation}
\label{mass}
M_{ej} \sim \frac{1}{2}\frac{\beta c}{\kappa} v_{ph} t_r^2 = 
   0.77 ~ M{_\odot} \left(\frac{\kappa}{0.1 {\rm ~cm^2~g^{-1}}}\right)^{-1} v_{ph,9} \left(\frac{t_r}{10 ~d}\right)^2, 
\end{equation}
where $\beta = 13.8$ is an integration constant, c the speed of light, $\kappa$ is the effective UVOIR opacity, 
$v_{ph}$ is the velocity at the photosphere with $v_{ph,9}$ in units of $10^9$ \cms, and $t_{r}$ the 
rise time to maximum light. This equation is based on assumptions of homologous expansion and 
self--similar diffusion with a power source at the center of spherical ejecta. Two further assumptions 
are that $v_{ph}$ is a reasonable proxy for the scaling velocity in the model and that the rise time to 
maximum light is a reasonable proxy for the effective timescale in the model, $t_{eff} = \sqrt{2t_d t_h}$, 
where $t_d$ is the model diffusion time and $t_h$ the model hydrodynamical time. In practice, the 
photospheric velocity will be affected by the distribution of density and opacity and hence may not 
precisely represent a fixed scaling velocity, the rise time is not exactly equal to $t_{eff}$ even in the 
context of the basic model (Chatzopoulos et al. 2013), and the opacity will be constant neither in 
space nor time.

The corresponding kinetic energy is then 
\begin{equation}
\label{ke}
E_{ke} = \frac{1}{2} M_{ej} <v^2>,
\end{equation}
which implicitly defines the mean squared expansion velocity. This velocity cannot be measured directly,
so a relation must be adopted between this velocity and the velocity at the photosphere. We adopt
the relation for a constant density sphere, $<v^2> = 3/5 v_{ph}^2$\footnote{Note that there is a typo
in Arnett (1982) where the relation between photospheric velocity, kinetic energy and mass is
given as $v_{ph}^2 = 3/5 (2E_{ke}/M_{ej})$. This error was corrected in Arnett (1996) where the
correct relation is given, $v_{ph}^2 = 5/3 (2E_{ke}/M_{ej})$, but the error has propagated in
the literature in, among other works, Valenti et al. (2008a), Chatzopoulos, Wheeler \& Vinko (2012)
and Chatzopoulos et al. (2013) who wrote $M_{ej} \sim \frac{3}{10}\frac{\beta c}{\kappa} v t_r^2$
without carefully specifying the prescription for the velocity.}.  With this relation we can then write
\begin{equation}
\label{ke2}
E_{ke} = \frac{3}{10} M_{ej} v_{ph}^2 =  \frac{3}{20} \frac{\beta c}{\kappa} v_{ph}^3 t_r^2 = 
    4.6\times10^{50} ~{\rm ergs} \left(\frac{\kappa}{0.1 {\rm ~cm^2~g^{-1}}}\right)^{-1} v_{ph,9}^3 \left(\frac{t_r}{10 ~d}\right)^2, 
\end{equation}
so $E_{ke}$ is proportional to $v_{ph}^3$ and hence sensitive to the choice of 
$v_{ph}$ and its dispersion among events. 

For the late-time tail, we adopt the formalism of Clocchiatti \& Wheeler (1997) and define a characteristic 
timescale
\begin{equation}
\label{T01}
T_0 = (\frac{C \kappa_{\gamma} M_{ej}^2}{E_{ke}})^{1/2}, 
\end{equation}
where C is a dimensionless structure constant dependent on the slope of the density profile and is typically 
C $\sim$ 0.05, and $\kappa_{\gamma}$  is the opacity to gamma rays. Using Equation (\ref{mass}) 
for $M_{ej}$, this can also be written 
\begin{equation}
\label{T02}
T_0 = \left[\left(\frac{5 \beta C}{3}\right)\left(\frac{c}{v}\right) \left(\frac{\kappa_{\gamma}}{\kappa}\right) \right]^{1/2}t_r =
      32 {\rm ~d} \left(\frac{\kappa_{\gamma}/0.03}{\kappa/0.1}\right)^{1/2} v_{ph,9}^{-1/2} \left(\frac{t_r}{10 ~d}\right), 
\end{equation}
where we take as a fiducial optical opacity $\kappa = 0.1$ cm$^2$ g$^{-1}$ and as a fiducial gamma--ray opacity
$\kappa_{\gamma} = 0.03$ cm$^2$ g$^{-1}$. 

\begin{figure}[htp]
\centering
\includegraphics[width=5 in, angle=0]{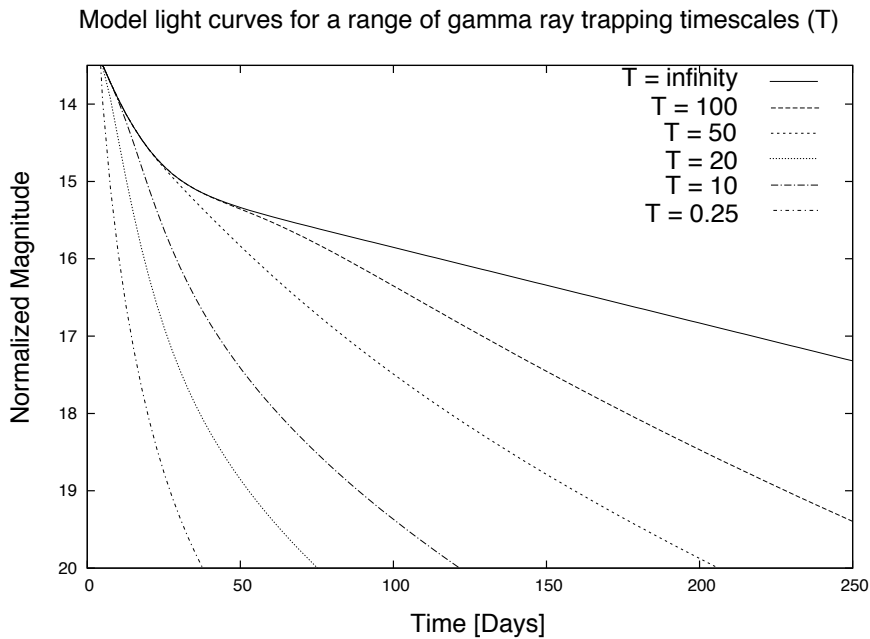}
\figcaption[models.ps]
{ Model light curves for a range of the gamma-ray trapping timescale $T_0$ (arbitrary normalization). A
trapping timescale of infinity corresponds to total trapping of gamma-rays from radioactive $^{56}$Ni
and $^{56}$Co. In these figures, positrons are assumed to have the same effective opacity as gamma-rays.
\label{models}}
\end{figure}

We take the radiated power to be 
\begin{equation}
\label{radiated}
L_{in} = L_{decay} \left(1 - e^{-(T_0/t)^2}\right),
\end{equation}
where L$_{decay}$ is the total input from gamma--rays and positrons from the
radioactive decay of $^{56}$Ni and $^{56}$Co according to 
\begin{equation}
\label{decay}
L_{decay} = A[e^{-t/t_{Ni}} + 0.21\times e^{-t/t_{Co}}],
\end{equation}
where A is a scaling constant that depends on the initial mass of $^{56}$Ni, $t_{Ni} = 8.8$ d is the 
decay time of $^{56}$Ni, and $t_{Co} = 111$ d is the decay time of $^{56}$Co. For this work, we neglect
the difference in the deposition functions of gamma-rays and positrons \citep{Capp97}, but comment 
on the separate effect of positrons below. Figure \ref{models} gives simple model bolometric light curves
with various degrees of gamma--ray leakage as determined by the timescale $T_0$. 

\section{Analysis}
\label{anal}

\begin{figure}[!htb]
\centering
\minipage{0.5\textwidth}
  \includegraphics[width=\linewidth]{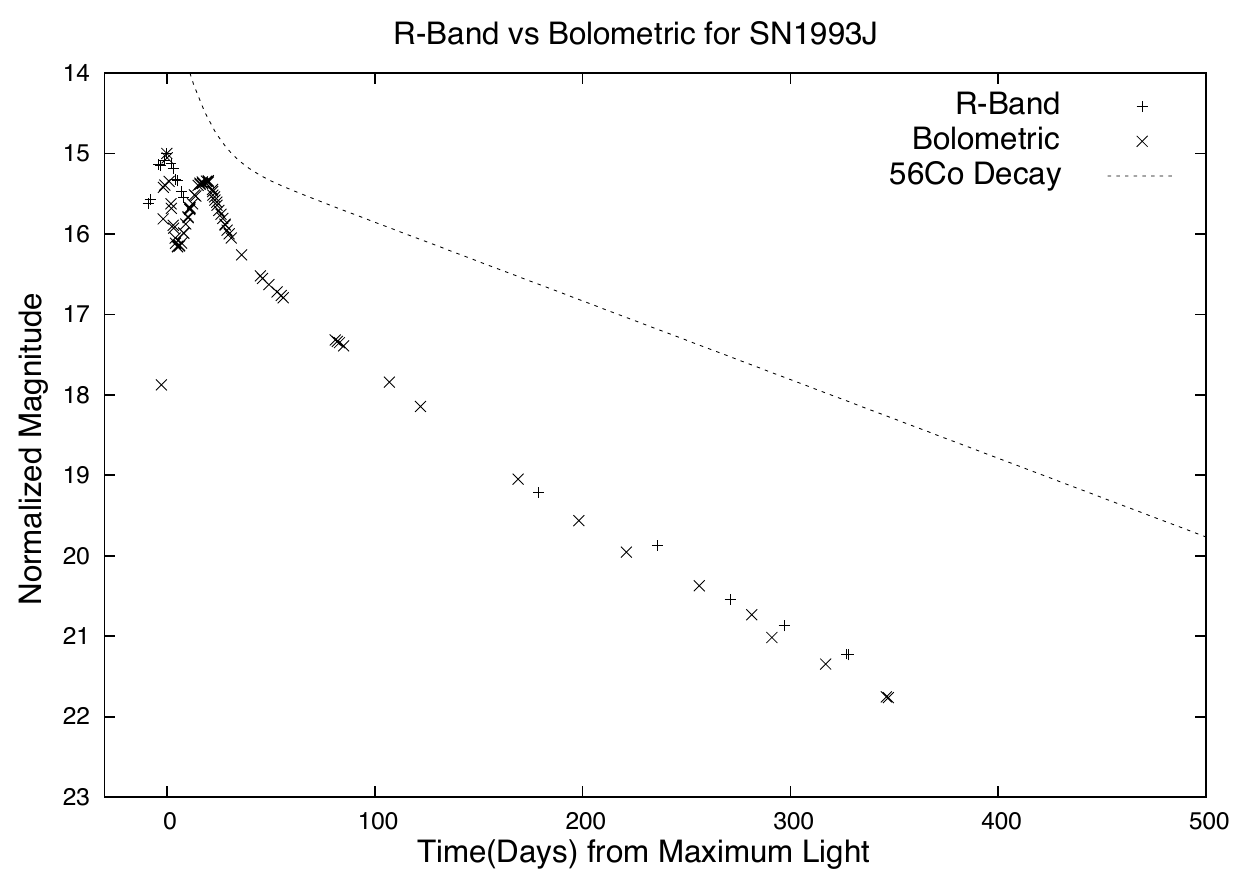}
\endminipage\hfill
\minipage{0.5\textwidth}%
  \includegraphics[width=\linewidth]{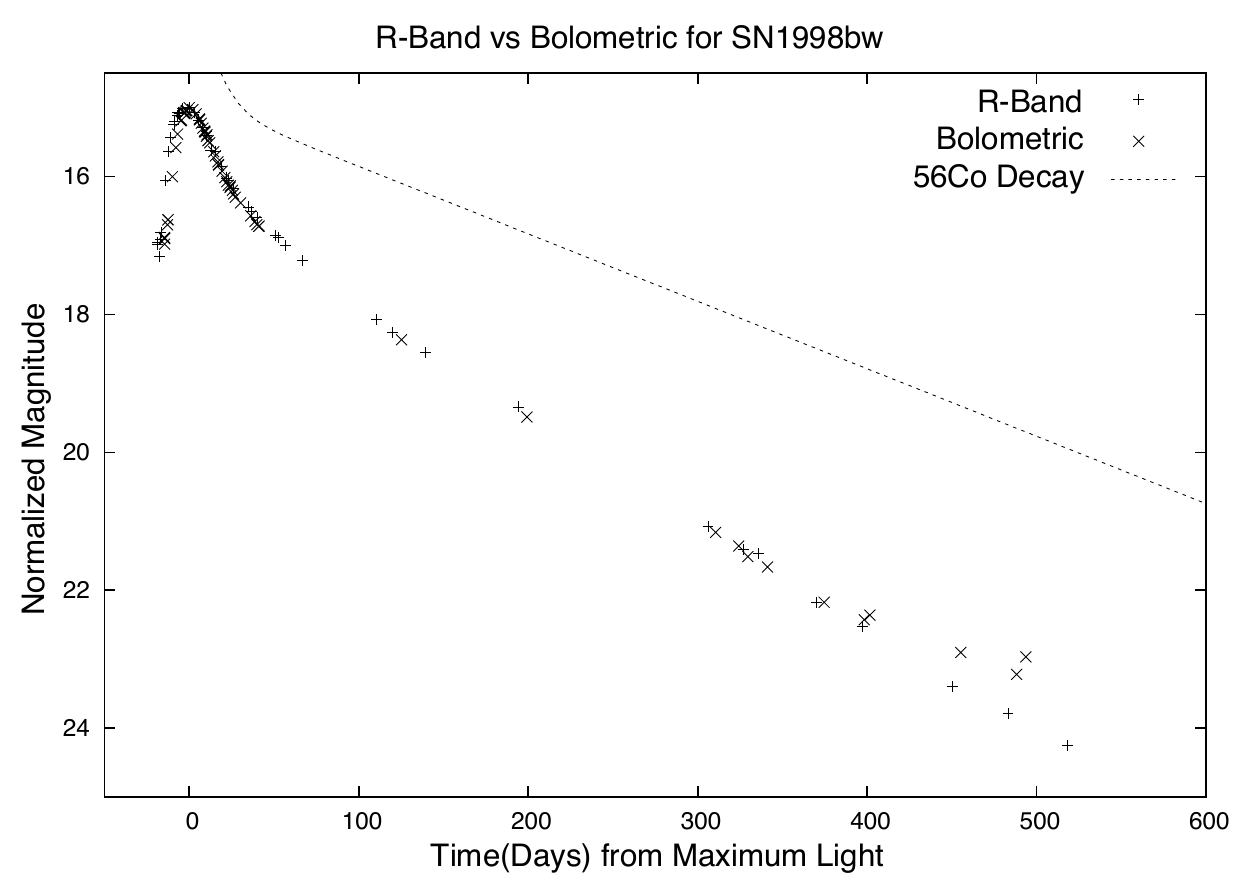}
\endminipage\hfill
\minipage{0.5\textwidth}
  \includegraphics[width=\linewidth]{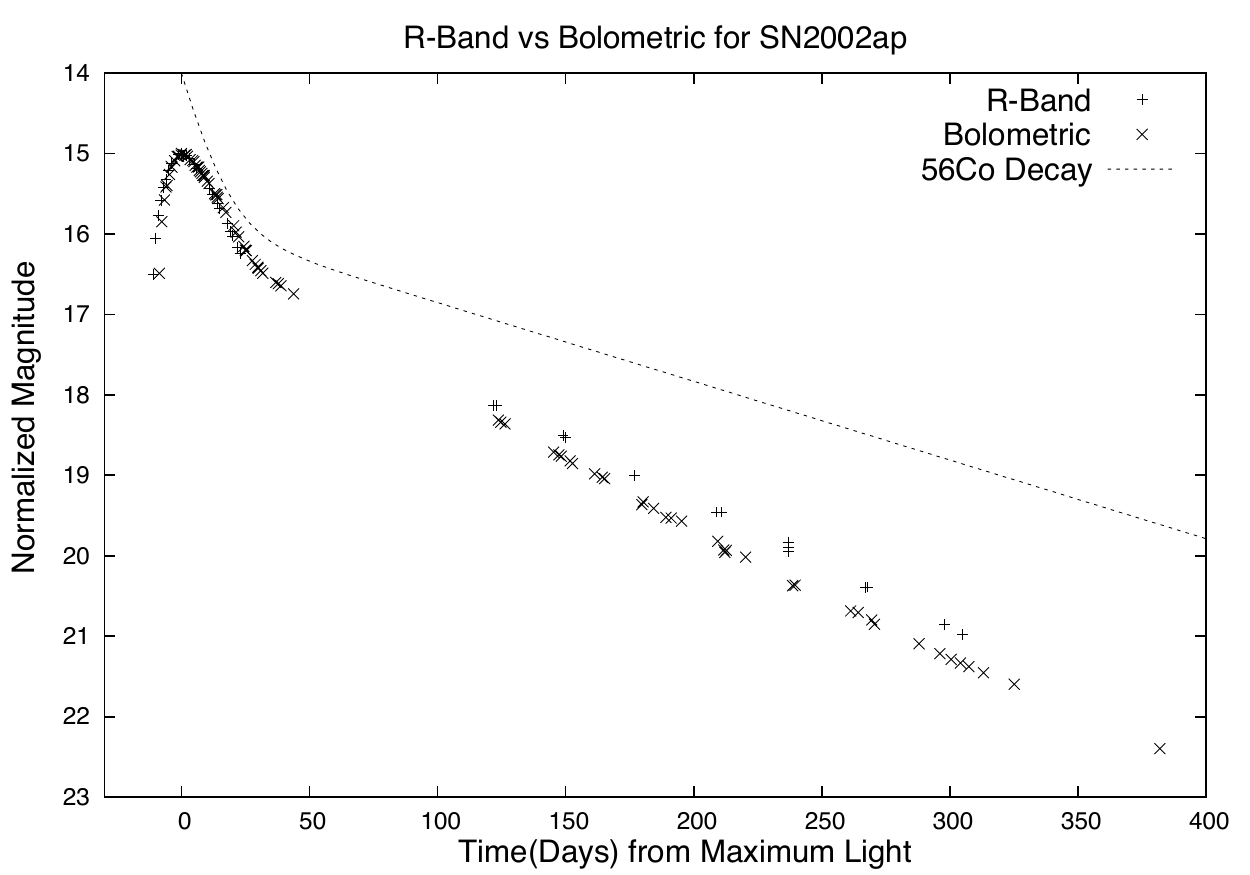}
\endminipage
\figcaption[BOL]
{Comparison of R--band and quasi--bolometric light curves for three supernovae: SN~1993J
(upper left panel); SN~1998bw (upper right panel); SN~2002ap (lower panel). Data are from 
\citet{B95,R94,Z04} (SN~1983J); \citet{P01,C11} (SN~1998bw); and \citet{Tom06} (SN~2002ap).
\label{RBC}}
\end{figure}

To estimate $M_{ej}$, $E_{ke}$, and $T_0$, we use data from the literature. Some of the supernovae that we 
discuss here have estimated UBVRI quasi--bolometric light curves, but not all. We have used the R--band 
as a proxy that allows us to treat most of the events in a somewhat homogeneous way, but also compare to
bolometric data where it is available. For events with only R--band data, we neglect bolometric corrections and 
assume that the R--band light curves are proxies of the bolometric luminosity to within a constant scaling 
factor (See Lyman, Bersier \& James 2014 for a discussion of estimating bolometric corrections for stripped--envelope
supernovae). Figure \ref{RBC} gives a comparison of the R--band and UVOIR light curves of three events, Type IIb 
SN~1993J, and two broad--lined SN~Ic, SN~1998bw, and SN~2002ap, normalized at peak light. This figure 
shows that, for these events, the R--band is a reasonable representation of the quasi--bolometric light
curve to times of hundreds of days. We have ignored reddening corrections for our R--band light curves, but 
such corrections have been made by the original authors who present the quasi--bolometric data. It seems 
unlikely that the results we present here are simply the result of variance in reddening, but this must be
borne in mind. In some cases of sparse or absent R--band data, we employ V--band or other available data. 

The sample for which there is good rise time data as well as long--time tail data is still sparse. The peak 
is used to estimate the rise time, $t_{r}$. For many of the more recent events and for some earlier 
nearby events, there is pre--maximum photometry. In the best cases, there are reasonably accurate 
estimates of the time of explosion (SN~1993J,  Richmond et al. 1994; SN~2008D, Malesani et al. 
2009, Modjaz et al. 2009). In other cases, the earliest data are up to two weeks prior to peak and thus are 
a good representation of the rise time (SN~1998bw, Patat et al. 2001; SN~1994I, Richman et al. 
1996; SN~1999dn, Benetti et al. 2011; SN 2007Y, Stritzinger et al. 2009; SN~2007gr, Valenti et al. 
2008a; SN~2009jf, Sahu et al. 2011; SN~2011bm, Valenti et al. 2012; SN~2011dh, Ergon et al. 2014, 
Marion et al. 2014). Note that even in these cases, the R--band peak may be delayed from the 
bolometric peak by several days. We have used the time to R--band peak in most cases, but 
recognize that some ambiguity is introduced in this way. 

In yet other cases there is some pre--maximum data, but not as extensive as the previous cases. For cases 
where there is pre--maximum data more than a magnitude dimmer than maximum, we use the time from first 
observation to the peak in R--band as an estimate for the rise time. This is an underestimate by perhaps 
several days, or typically 10 -- 20\%. Examples are SN~1983N (FES data; Clocchiatti et al. 1997), 
SN~1996cb (Qui et al. 1999), SN~2002ap (Tomita et al. 2006), SN~2004aw (Taubenberger et al. 2006), and
SN~2009bb (Pignata et al. 2011). 

In other cases, there is little or even no pre--maximum data. In these cases, we make estimates based 
on analogies to other events (SN~1983V, Clocchiatti et al. 1997; SN~1984L, Swartz \& Wheeler 1991) 
or on the decline from maximum and estimate the time to decline by a magnitude from peak (SN 1990B, 
Clocchiatti et al. 2001). The light curves of the stripped--envelope  events are typically asymmetric around 
the peak with more rapid rise than decline, so this post--maximum decline timescale almost surely 
overestimates the rise time. In these various cases, we have put our estimates in square brackets in 
Table \ref{tab:timescale} to highlight the uncertainty. The rise time comes in squared in the estimates 
of the mass and energy, so these estimates should be considered with caution. 

Taddia et al. (2014) give a table containing observed R--band rise times. Most of these agree with our choices,
but there are some discrepancies. We chose 17 d for the rise for SN~1993J versus 22.53 d
for Taddia et al.  Our estimate is measured from the early minimum in the light curve, ignoring the initial fireball 
decline phase. We assume the subsequent rise is more representative of other events that were not
caught so early and of the diffusion time we are attempting to constrain. We chose 15 d for SN~2007Y
from Stritzinger et al. (2009), versus the 21 d assigned by Taddia et al. The latter seems to involve
some extrapolation even beyond the early V--band data, but our value is surely a lower limit.

Estimates of velocity are taken from photospheric velocities given in the literature. For uniformity, we 
have tried to select whenever possible the velocity of the P Cygni absorption minimum of Fe II
$\lambda$5169 measured at R--band maximum. While this line may be relatively free of blending, as may affect
estimates based on Si II, there are still issues. In particular, the velocity measured in this way often
declines rather rapidly from pre--maximum to post--maximum epochs. Velocities at bolometric 
maximum several days earlier would usually have been somewhat higher and there is the general
concern as to whether any velocity measured in this way properly represents the scale velocity 
of the underlying model. Such considerations comes in especially sensitively in estimates of the energy.
Cano (2013) and Lyman et al. (2014) also give compilations of photospheric velocities. In general, our 
estimates agree with theirs to within 20\%. Exceptions are SN~2004aw where we determined 16,000 \kms, 
Cano gave 11,800 \kms\ and Lyman et al. give 11,000 \kms, and SN~2009bb, for which we determined 
18,000 \kms, Cano gave 15,000 \kms, and Lyman et al. gave 17,000 \kms.

We use the late--time light curves to directly determine an estimate of $T_0$ and one 
standard deviation error bars by minimizing the $\chi^2$ of the fit of Equation \ref{radiated} to 
the data. To minimize the effect of scatter in the peak data, we only use the data after 50 d 
from maximum to determine the observed value of $T_0$. Table \ref{tab:timescale} also gives 
the measured values of $T_0$. The error given for SN~2007Y is especially, and probably artificially, 
low because there is sparse late--time data that is easy to fit. 

For some objects in the sample here, Clocchiatti et al. (2008) also estimated values of $T_0$ from 
UBVRI quasi--bolometric light curves. Comparing the values derived here (first numbers, R--band; 
the numbers in parentheses are our estimate of the value based on quasi--bolometric data) with 
those given by Clocchiatti et al. (second numbers) we find: SN~1994I, 115 d vs 65 d; SN~1998bw, 
192 (283) d vs 120 d; SN~2002ap, 186 (202) d vs 142 d; SN~1990B, 113 d vs 88 d. There are 
differences of 10 to 30\% with our estimates somewhat on the high side. The difference is most
likely due to our simple treatment that implicitly has the trapping of positrons decreasing in the
same manner as for the gamma--rays. If positrons are trapped, as assumed by Clocchiatti et al.,
then for values of $T_0 \sim$ 100 to 200d, the positron contribution will be comparable to that
of the gamma--rays by 200 -- 300 d. With this extra contribution to the luminosity, a light curve 
of given late--time slope can be accommodated with a smaller value of the parameter $T_0$. In 
the first portion of our discussion comparing the observed decay time versus that predicted from 
the peak properties, as expressed in Equation \ref{T02}, differences of $\la 30$\% are not significant
since we are interested in discrepancies of factors of several. In \S \ref{discuss} where we address 
the issue of estimating ejecta masses and energies from the tail, the decay time enters as the 
square and hence more sensitively. We address the issue of positron trapping again there.

\begin{figure}[htp]
\centering
\includegraphics[width=5 in, angle=0]{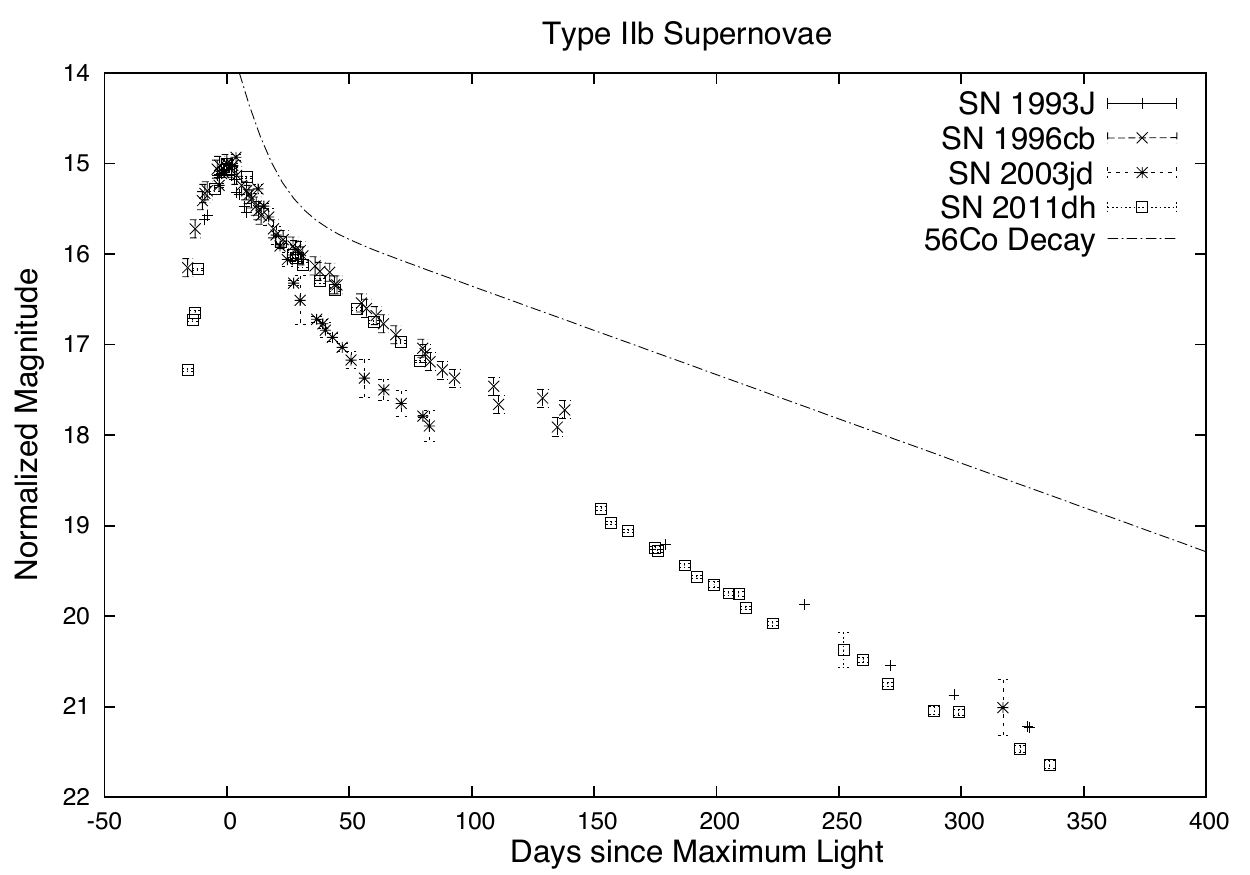}
\figcaption[snIIb.ps]
{R--band light curves of a sample of SN~IIb.  Note the similar behavior around maximum and the 
very similar late--time light curves of SN~1993J and SN~2011dh for nearly 250 d, but the 
variation in the decline rate manifested by SN~ 1996cb (slower) and SN~2003jd (faster after
peak, but comparable to SN~1993J later). The dot--dash line represents the total trapping of the 
energy of radioactive decay. Data are from \citet{B95,R94,Z04} (SN~1993J); \citet{Q94} (SN~1996cb); 
\citet{V08b} (SN~2003jd); \citet{Tsv12} (SN~2011dh; note that even later data is given in Ergon et al. 2015).
\label{snIIb}}
\end{figure}

\begin{figure}[htp]
\centering
\includegraphics[width=5 in, angle=0]{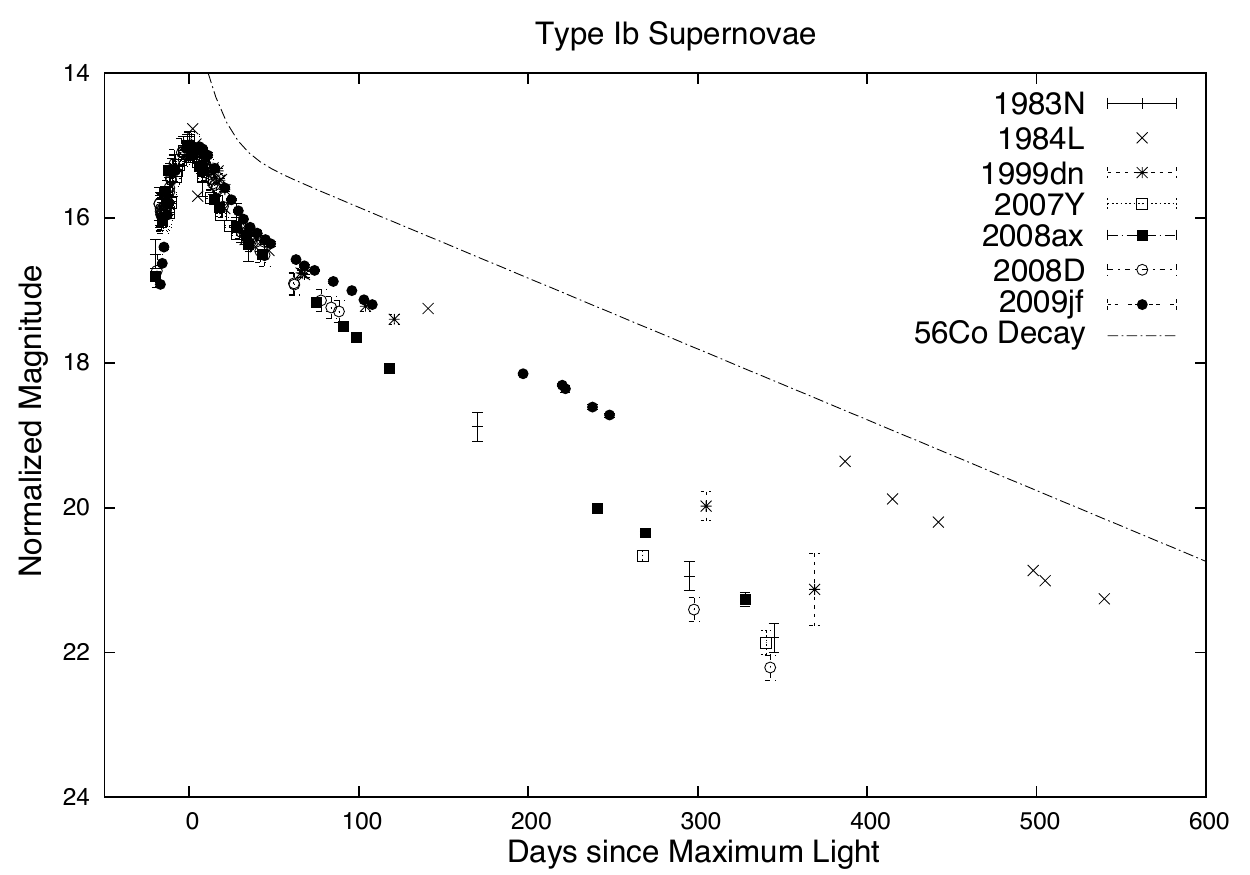}
\figcaption[snIb.ps]
{R--band light curves of a sample of SN~Ib. Note the significant dispersion in the decline rate
at late time. The dot--dash line represents the total trapping of  the energy of radioactive decay.
Data are from \citet{C96} (SN~1983N); \citet{SW91} (SN~1984L); \citet{B11} (SN~1999dn); 
\citet{S09} (SN~2007Y); Malesani et al. (2009), Modjaz et al. (2009) (SN~2008D); 
\citet{Pasto08} (SN~2008ax); \citet{S11} (SN~2009jf). 
\label{snIb}}
\end{figure}

\begin{figure}[htp]
\centering
\includegraphics[width=5 in, angle=0]{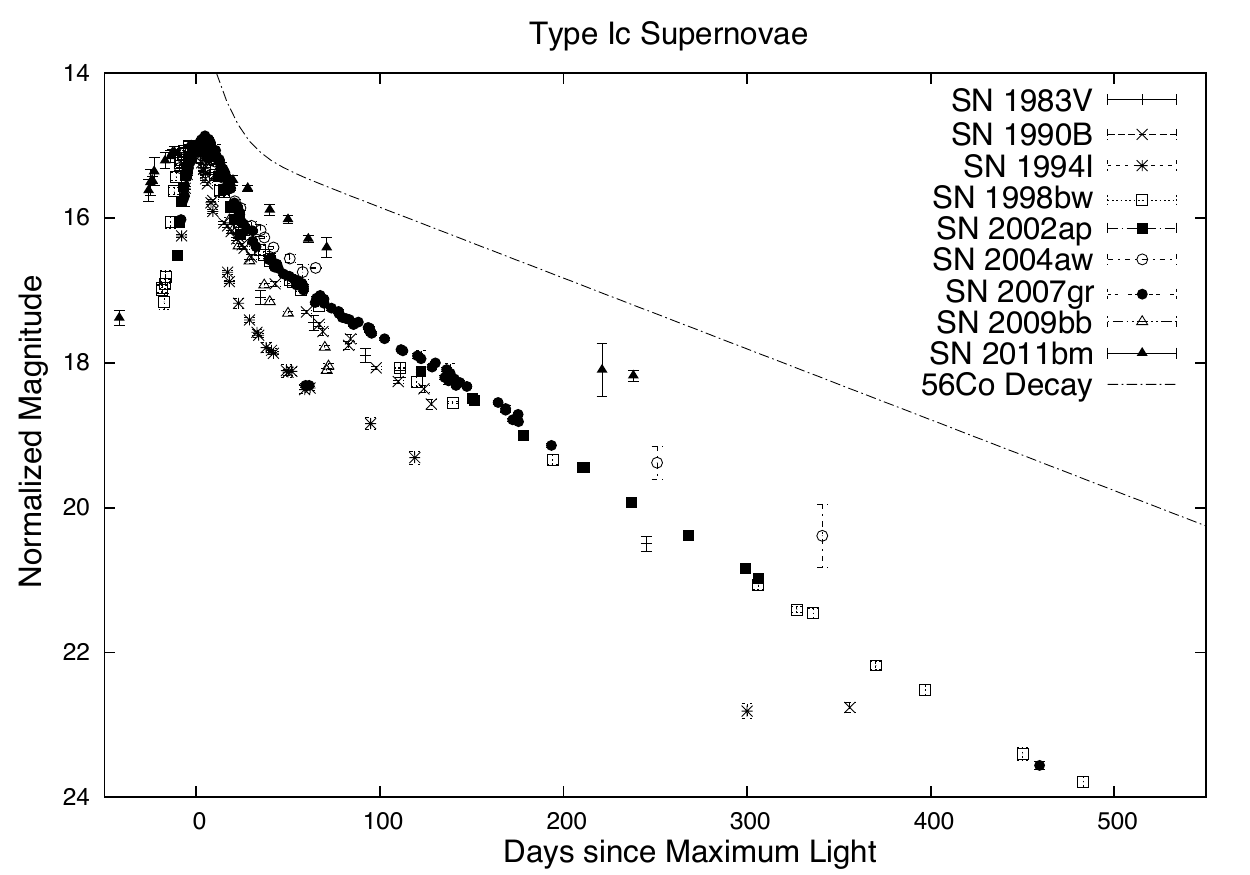}
\figcaption[snIc.ps]
{R--band light curves of a sample of SN~Ic. The data for SN 1983V are V--band. Note the distinct rapid rise 
and decline of SN~1994I and the slower rise and decline of SN~2011bm. The post--peak decay of SN~2011bm 
closely follows the radioactive decay line. Otherwise, the relative homogeneity of the peaks leads to a large 
dispersion of late--time tails. The dot--dash line represents the total trapping of the energy of radioactive decay.
Data are from \citet{C97} (SN~1983V); \citet{C01} (SN~1990B); \citet{C08} (SN~1994I); \citet{P01} (SN~1998bw);
\citet{Tom06} (SN~2002ap); \citet{Taub06} (SN~2004aw); \citet{P11} (SN~2009bb); \citet{V12} (SN~2011bm).
\label{snIc}}
\end{figure}

Figures \ref{snIIb},  \ref{snIb}, and \ref{snIc} give the light curves we have compiled. For comparison 
purposes, the light curves have been shifted in time to a common peak and have been normalized
to an arbitrary magnitude of 15.

Figure \ref{snIIb} gives a sample of 4 SN~IIb. They are all rather similar near peak, with similar rise times
and early declines, but substantial departures from homogeneity set in beginning about 30 d after
maximum. The long--term light curves of these events are distinctly steeper than that given by full trapping of  
gamma-rays from \co\ decay. There is a clear dispersion of the late--time decay times despite the similarity of 
the light curves near peak and the common spectral class. SN~1993J and SN~2011dh show very similar 
late--time light curves for nearly 350 d. SN~2003jd falls more steeply around 30 d after peak, but 
has a similar rate of decline at later times. SN~1996cb is slower at late time than the others. Its decay after 
100 d is similar to that of total trapping, but a contribution from circumstellar interaction cannot be ruled out.
 
Figure \ref{snIb} gives a sample of 7 SN~Ib. We have plotted V--band data for SN~1983N, r`--band data for 
SN~2007Y, and bolometric data for SN~2008D and SN~2008ax. SN~2008ax is often classified as a SN~IIb 
because some hydrogen was seen early on. By maximum, it looked like a classic SN~Ib. We have classified
it as a SN~Ib here, but that particular choice does not affect our analysis in any substantial way. These seven 
events again are rather similar on the rise. SN~1984L flattens to be comparable to the radioactive decay 
slope, but SN~1984L was a late--time radio emitter, so there might be some contribution to the light curve 
from collision with a circumstellar medium. SN~1999dn and SN~2009jf are similar on the tail for the first 
100 d. SN~2009jf subsequently falls at a rate roughly comparable to cobalt decay until at least 250 d, 
but SN~1999dn falls more steeply at 300 to 400 d. SN~1983N, SN~2007Y, SN~2008D and 
SN~2008ax fall at very similar rates at late times, with SN~2008D being the steepest of this group.

Figure \ref{snIc}  gives a sample of 9 SN~Ic. We have plotted B--band data for SN~1983V. There appears to be 
some true dispersion in the early peak widths. SN~2011bm has a distinctly wider peak. SN~1994I falls especially 
rapidly after maximum. SN~2009bb nearly rivals SN~1994I in terms of steepness of immediate post--maximum 
decline. There is clearly a large dispersion in the rate of decay of the late--time tails of the SN~Ic. SN~2011bm 
continues to decline slowly, with a late--time decay very comparable to that expected for \co\ decay. The late--time 
light curves of the broad--line Type Ic SN~1998bw and SN~2002ap are very similar; SN~2004aw declines 
somewhat more slowly than the two broad--line events, and SN~1990B somewhat more rapidly. The light curve 
of SN~1994I flattens at very late times (300 d), which may have to do with partial trapping of positrons 
(Clocchiatti et al. 2008). 

\begin{figure}[htp]
\centering
\includegraphics[width=5 in, angle=0]{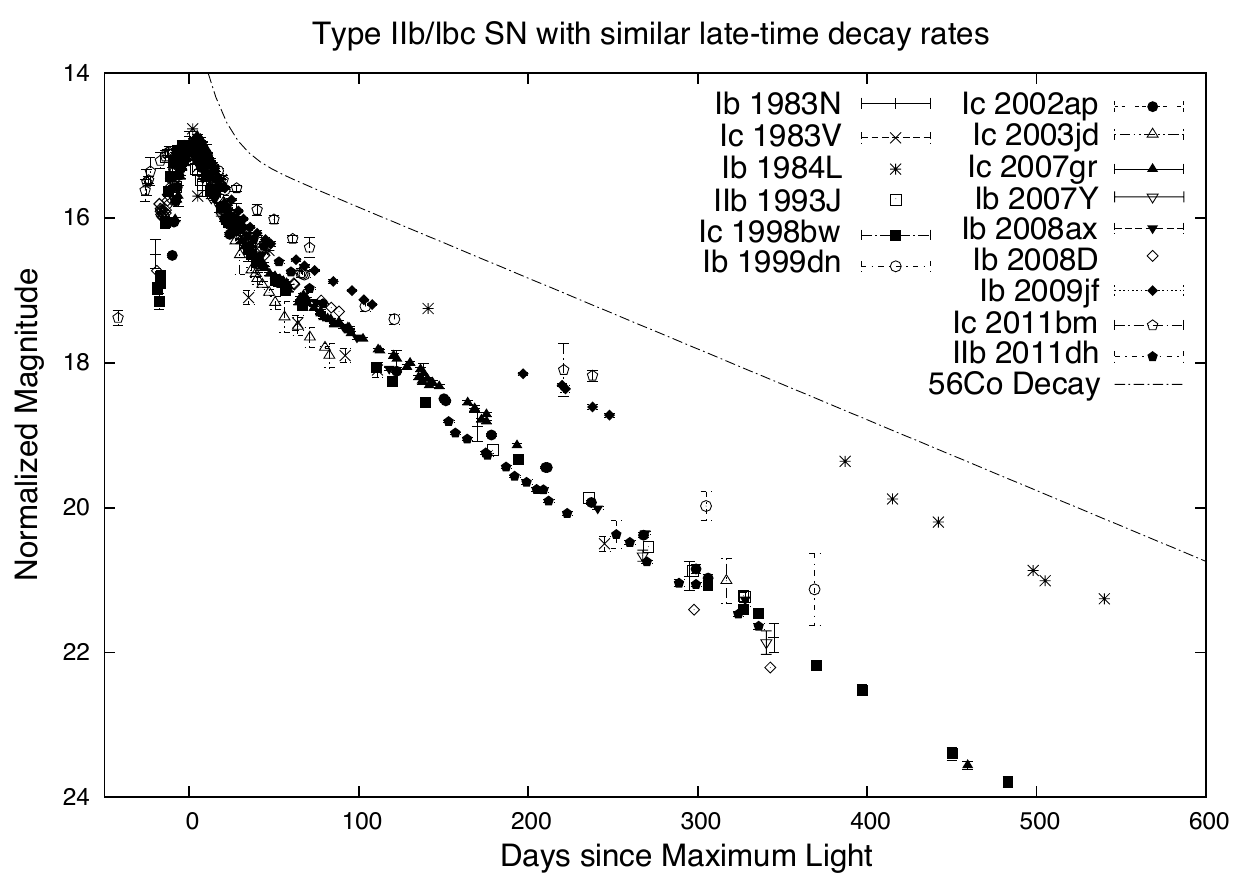}
\figcaption[similar.ps]
{Light curves of a selected sample of SN~IIb, Ib, and Ic with similar late--time decay rates. Note 
the relative homogeneity of the peaks and the bifurcation (in this small sample) of the late--time 
light curves into two groups. The slower--declining group has only SN~Ib, but the 
more rapidly--declining group contains SN~IIb, SN~Ib and SN~Ic, including high--velocity SN~Ic. 
The dot--dash line represents the total trapping of  the energy of radioactive decay.  
References to data are given in previous captions.
\label{similar}}
\end{figure}

Figure \ref{similar} gives a selected sample of light curves that seem, by eye, to fall into two categories 
in terms of late--time tails. One group has a decline roughly comparable to \co\ decay, and the other is 
significantly steeper. Among the first category with rather shallow late--time declines, Type Ib SN~1984L, 
Type Ic SN~2011bm, and perhaps Type Ib SN~2009jf are the only examples in our sample of late-time 
light curves that track \co\ decay (corresponding to very large values of $T_0$). As remarked above, 
the late--time light curves of the Type Ib SN~1999dn and SN~2009jf are very similar up to 100 d, 
with SN~1999dn perhaps tracking \co\ decay. Later data on SN~1999dn fall below this early common trend. 
Either SN~1999dn was not powered by \co\ decay earlier, but perhaps by circumstellar interaction, 
or some other effect, perhaps dust formation, arose after 200 d to suppress the optical light curve.
In any case, most stripped--envelope supernovae decay considerably faster than expected for complete 
trapping of \co. 

\begin{figure}[htp]
\centering
\includegraphics[width=5 in, angle=0]{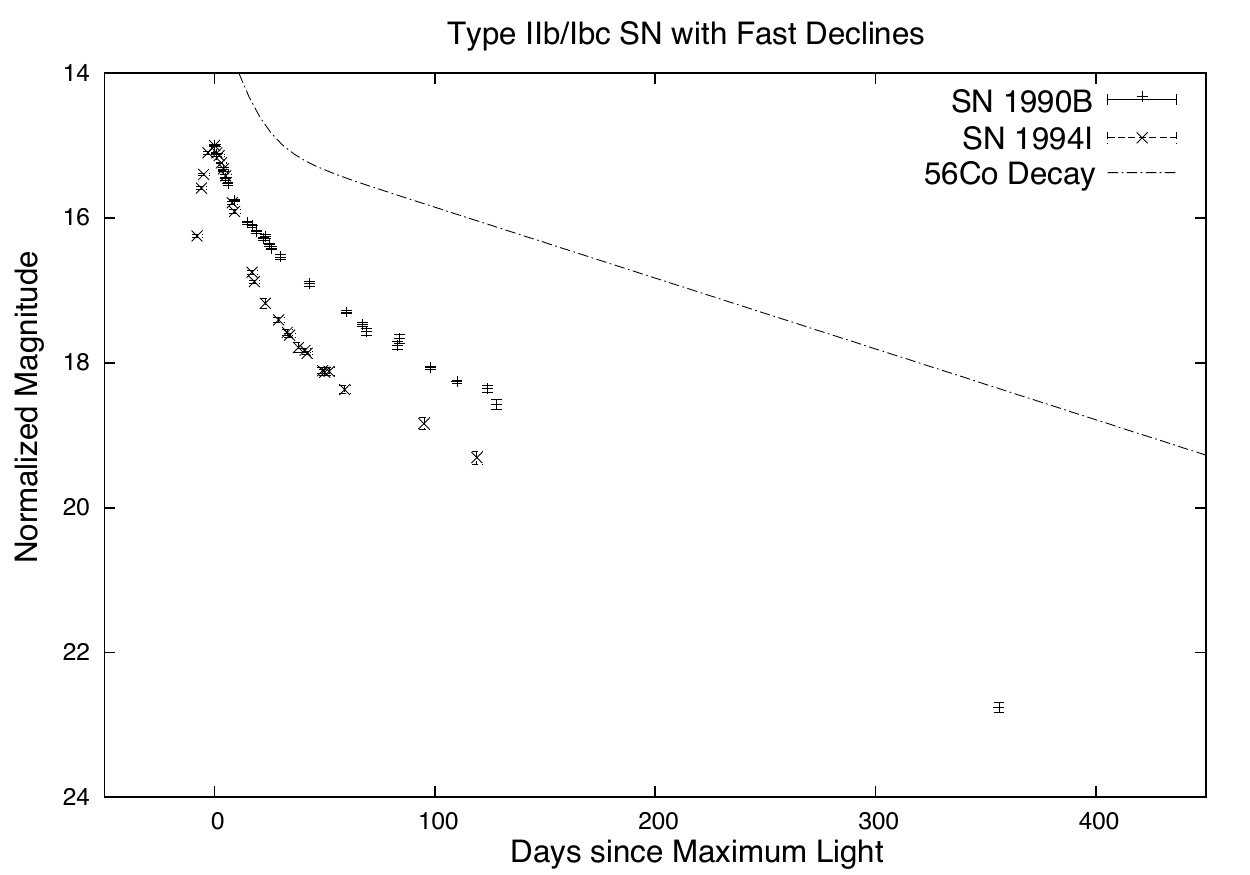}
\figcaption[fast.ps]
{Light curves of two stripped--envelope supernovae with rather steep late--time decay 
rates, both SN~Ic. The dot--dash line represents the total trapping of  the energy of 
radioactive decay.  References to data are given in previous captions.
\label{fast}}
\end{figure}

The assortment of light curves of other SN~IIb, SN~Ib, and SN~Ic, including some broad--line SN~Ic, 
that fall in the second category in Figure \ref{similar} are also rather similar to one another, but with 
a distinctly different, steeper slope than those that might follow \co\ decay. Interestingly, this group 
comprises 11 of our sample of 20 events. This group includes SN~1983V, but we note that the data 
are B--band, so the decline rate may be exaggerated. The light curves of SN~1983V in both V--band 
and B--band were very similar to those of SN~1993J in the respective bands (Clocchiatti et al. 1997), 
and the light curve we plot here is very similar to that of SN~1993J in R--band at late times, but it is 
appropriate to regard this comparison with some caution. There appears to be some dispersion in 
peak width for this particular subsample. Figure \ref{fast} presents data for two events that decay more 
rapidly at late times than most. Both are SN~Ic. SN~1994I seems to fall more rapidly from peak than 
SN~1990B.

Figures \ref{similar} and \ref{fast} illustrate that neither spectral type nor similar peak light curves determine the 
slope of the late--time tail, and that events of different spectral type can have remarkably similar 
late--time tails, as emphasized by Clocchiatti et al. (1996, 1997). The spectra and light curves of 
stripped--envelope supernovae near maximum light are not sufficient to fully classify them.

\begin{center}
\begin{deluxetable}{lcccccc}
\tabletypesize{\footnotesize}
\tablewidth{0pt}
\tablecaption{Estimated vs. Observed Timescale of Late--Time Decay of Stripped--Envelope Supernovae \label{tab:timescale}}
\tablecolumns{7}
\tablehead{
\colhead{Event} & 
\colhead{Type} &  
\colhead{$v_{ph}$} & 
\colhead{$t_{rise}$} & 
\colhead{$T_{0}$} & 
\colhead{$T_{0}$} & 
\colhead{Band} \\ 
\colhead{SN}  &
\colhead{}  & 
\colhead{$10^{9} cm s^{-1}$} & 
\colhead{d} & 
\colhead{Est.(d)} &  
\colhead{Obs.(d)}  &
\colhead{}
}
\startdata
    1983N    & Ib     & $0.78\pm0.08$           & $[16\pm3]$$^*$           & $58\pm12$        & $ 166\pm16 $    & V        \\
    1983V    & Ic      & $1.4\pm0.14$             & $[20\pm4]$$^*$           & $54\pm11$        & $151\pm7   $    & B     \\
    1984L    & Ib      & $1.0\pm0.1$              & $[20\pm4]$$^*$             & $64\pm13$        & $660\pm109$     & V   \\
    1990B    & Ic      & $1.1\pm0.1$              & $[20\pm4]$$^*$            & $61\pm13$        & $113\pm5$      & R          \\
    1993J     & IIb     & $0.75\pm0.08$          & $17\pm1.7$                 & $63\pm7$          & $135\pm5$     & Bol       \\
    1994I     & Ic      & $1.15\pm0.12$           & $9\pm0.9$                   & $27\pm3$          & $115\pm6$     & R      \\
    1996cb  & IIb     & $0.9\pm0.1$               & $25\pm2.5$                 & $85\pm10$        & $152\pm15$     & R     \\
    1998bw  & Ic      & $1.4\pm0.14$            & $14.4\pm1.4$              & $39\pm4$        & $192\pm6$     & R      \\  
    1999dn  & Ib      & $1.0\pm0.1$               & $[15\pm3]$$^*$           & $48\pm10$        & $220\pm14$     & R      \\
    2002ap  & Ic       & $1.5\pm0.15$            & $[9\pm0.18]$$^*$        & $24\pm5$          & $186\pm8$    & Bol     \\
    2003jd   & IIb     & $1.4\pm0.14$             & $[20\pm4]$$^*$           & $54\pm11$         & $237\pm81$       & R    \\
    2004aw  & Ic      & $1.6\pm0.16$            & $[35\pm7]$$^*$            & $89\pm18$         & $206\pm123$      & R     \\
    2007Y    & Ib      & $0.9\pm0.09$             & $15\pm1.5$                & $51\pm6$           & $114\pm1$        & r'     \\
    2007gr   & Ic      & $0.64\pm0.06$           & $15\pm1.5$                 & $60\pm7$          & $157\pm4$     & R\\
    2008D    & Ib      & $1.0\pm0.1$              & $20\pm2$                    & $64\pm7$           & $107\pm3$         & R\\
    2008ax   & Ib      & $0.8\pm0.08$            & $22\pm2.2$                 & $79\pm9$           & $175\pm11$   & Bol\\
    2009bb   & Ic      & $1.8\pm0.18$            & $[9\pm1.8]$$^*$          & $22\pm4$           & $57\pm20$      & R    \\
    2009jf     & Ib      & $1.0\pm0.1$              & $25\pm2.5$                 & $80\pm9$           & $315\pm24$    & R      \\
    2011bm  & Ic      & $0.75\pm0.08$           & $45\pm4.5$                & $170\pm19$        & $308\pm41$      & R       \\
    2011dh   & IIb     & $0.69\pm0.07$           & $22\pm2.2$                & $85\pm10$          & $210\pm10$       & R      \\
\enddata
\vspace{1mm}
$^*$\footnotesize{Entries in square brackets are especially uncertain; see text}
\label{tab:timescale}
\end{deluxetable}
\end{center} 

\begin{figure}[htp]
\centering
\includegraphics[width=5 in, angle=0]{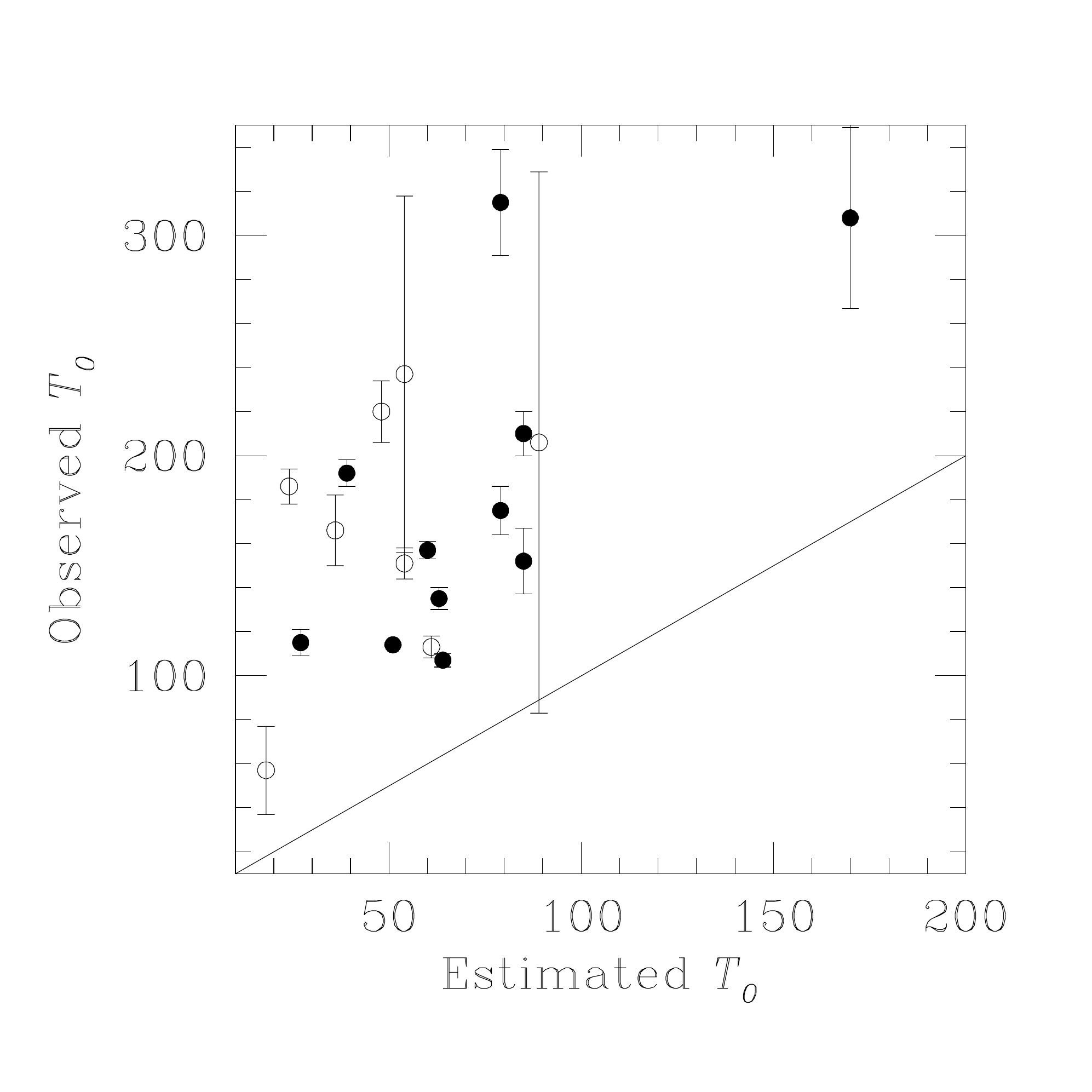}
\figcaption[compare.ps]
{The values of the timescale, $T_0$, obtained from fits to the late--time data
(with 1--$\sigma$ error bars) are plotted against the values of $T_0$ predicted
from the properties of the peak according to Equation \ref{T02}. The diagonal
line represents equality of the two time scales. The observed values are 
considerably higher than the predicted values, indicating a significantly slower
late--time decay than predicted. Open symbols represent events for which
the rise time is especially uncertain (square brackets in Table \ref{tab:timescale}). 
Data for SN~1984L have been omitted as they are off the scale of the plot.
\label{compare}}
\end{figure}

Table \ref{tab:timescale} gives for our sample of events, the observed values of $v_{ph}$, $t_{r}$, 
and $T_0$ and the value of $T_0$ estimated from Equation \ref{T02}. We have also attempted 
to assign representative uncertainties to our estimates of $v_{ph}$ and $t_{r}$. For $v_{ph}$ 
we have assigned an uncertainty of 10\%, or about $\pm1000$ \kms\ for the typical observed 
velocities. While not exact, this is a representative uncertainty. For $t_{r}$, we have also assigned 
an uncertainty of 10\% for many of the events, 2 or 3 d for typical events, but an uncertainty 
of 20\% for the events where, for various reasons, the rise time is especially ill--determined. 
The values of $t_{r}$ for these events are given in square brackets in Table \ref{tab:timescale} 
to draw attention to these events. We have then propagated these uncertainties through the 
estimate of $T_0$ from Equation \ref{T02} and for other quantities to be discussed in \S 
\ref{discuss}. We have assumed for this exercise that the observed values of $v_{ph}$, $t_{r}$, 
and $T_0$ are uncorrelated, but this may not be completely true. The predicted value of $T_0$ 
in Equation \ref{T02} only depends linearly on the rise time and on the square root of the velocity. 
Uncertainty in these parameters are relatively unimportant to that particular quantity. The effects 
in which we are interested are typically factors of several in $T_0$, as illustrated in 
Table \ref{tab:timescale}.

For illustration, we have also made an analytic fit to the tail for the group of 11 events that form a 
nearly common locus in Figure \ref{similar}. Restricting the analysis to more than 50 d from 
maximum, we can neglect the contribution of the $e^{-t/t_{Ni}}$ term in Equation (\ref{decay}). 
Anticipating that $T_0$ is fairly large, then at 50 d the term $e^{-(T_0/t)^2}$ in Equation 
(\ref{radiated}) is essentially zero and can also be neglected. Making those two assumptions, 
one can write the magnitude $M = -2.5log L + K$ at the two epochs, 50 d and 400 d, 
where the similar track in Figure \ref{similar} passes through about magnitude 17 and 22.5, 
respectively. Using these values, the resulting parameter is $T_0 = 158$ d, corresponding 
to $\sim 0.016$~\magd, a reasonable estimate to the range of values for these 11 events presented 
more formally in Table \ref{tab:timescale}.

Figure \ref{compare} shows the observed value of $T_0$ versus that estimated from the 
peak properties using Equation \ref{T02}. This exercise shows clearly that the values of 
$v_{ph}$ and $t_r$, and hence the values of $M_{ej}$ and $E_{ke}$, derived from the peak 
behavior are poor predictors of the late--time behavior. The discrepancies are not subtle. The 
directly measured values of $T_0$ occasionally agree within a factor of two of those estimated 
from the peak properties, but the discrepancies are routinely factors of 2 to 4, and can be as 
much as a factor of 10(SN~1984L). In no case is the estimate of $T_0$ from the peak properties 
greater than that determined from direct observations of the late--time tail.

\section{Discussion}
\label{discuss}

This analysis defines four dilemmas associated with the late--time light curves
of stripped--envelope supernovae. (1) The relatively similar peak properties often
belie a considerable spread in the properties of the late--time tail. (2) Even for
events with similar peaks and tails, the properties derived from the peak seem to be
incommensurate with the behavior of the tail. (3) Events with different spectral 
types can, nevertheless, have rather similar late--time decay. (4) Both typical
SN~Ic and broad--lined SN~Ic can have similar late--time decay despite the
nominal significant differences in energy associated with the broad lines of
the latter. 

To elucidate these various, related, dilemmas, we can write expressions for the
ejected mass and energy in terms of the observed properties as follows. From
the peak, we have from Equations \ref{mass} and \ref{ke2}
\begin{equation}
\label{masspeak}
M_{ej} = \frac{1}{2}\frac{\beta c}{\kappa} v_{ph} t_r^2, 
\end{equation}
and
\begin{equation}
\label{kepeak}
E_{ke} =  \frac{3}{20} \frac{\beta c}{\kappa} v_{ph}^3 t_r^2, 
\end{equation}
and from the tail we have from Equation \ref{T01}, invoking $E_{ke} = \frac{3}{10} M_{ej} v_{ph}^2$,
\begin{equation}
\label{masstail}
M_{ej} = \frac{3}{10}\frac{v_{ph}^2 T_0^2}{C \kappa_{\gamma}}, 
\end{equation}
and
\begin{equation}
\label{ketail}
E_{ke} =  \left(\frac{3}{10}\right)^2 \frac{v_{ph}^4 T_0^2}{C \kappa_{\gamma}}. 
\end{equation}

The complex relationships among the early and late photometric behaviors of different 
spectral types of stripped--envelope supernovae can be illustrated by comparing light 
curves of several well--observed events that have data in Table \ref{tab:timescale}. 
The two Type~Ib SN~1994L and SN~2008D have very similar rise times and photospheric
velocities, yet very different late--time tails, illustrating point (1). The typical Type~IIb event 
SN~2011dh and the hydrogen--contaminated Type~Ib event SN~2008ax had similar rise 
times and photometric velocities of $\sim 22$~d and $\sim 8,000$~\kms, respectively. 
For similar opacities, one might expect similar ejecta masses and energies and thus similar 
tails. The tails were similar, but also similar to that of Type Ic SN~2007gr, a congruence 
that defies easy explanation, point (3) above. 

Furthermore, not all events with similar tails have similar peak properties. SN~2007gr 
had a somewhat more rapid rise, $\sim 15$~d, and lower photospheric velocity near peak, 
$\sim 6400$~\kms. Both of these differences suggest that, for similar opacities, 
the ejecta mass ($\propto v_{ph} t_r^2$; Equation \ref{masspeak}) and the 
kinetic energy ($\propto v_{ph}^3 t_r^2$; Equation \ref{kepeak}) should be 
less than for SN~2011dh and SN~2008ax. The fact that the late--time tail of 
SN~2007gr is similar to those of SN~2011dh and SN~2008ax and others of different 
spectral class would then seem to require an unexpected correlation of the ejecta 
mass and energy. The broad--line Type~Ic SN~2002ap had a rise time of $\sim 10$ d 
and a photospheric velocity of $\sim 15,000$~\kms. Based on the peak properties, 
and again assuming similar opacities, SN~2002ap should have about the same ejecta 
mass as SN~2007gr, but a kinetic energy that is larger by about a factor of 4. It is difficult 
in this simple framework to understand how their late--time tails are similar, illustrating
point (4).

Equations \ref{masstail} and \ref{ketail} show that the constraint of similar late--time light 
curves, those with similar values of $T_0$, implies that 
$M_{ej} \propto v_{ph}^2$ and $E \propto M_{ej}^2 \propto  v_{ph}^4$. With 
these relations, the similar tails imply that the ejecta mass of SN~2002ap must be 
larger than that of SN~2011dh and SN~2008ax by a factor of $\sim 4$ and 
the energy larger by a factor of $\sim 16$ while maintaining similar late--time 
decay.  

Even for individual events with similar peak and tail properties, it is difficult to reconcile the 
two phases, point (2) above. The constraints from the width of the peak, $M_{ej} 
\propto v$ and $E \propto v^3$ for a constant opacity and given light curve rise time, 
are different from those associated with the slope of the tail, $M_{ej} \propto v^2$ and 
$E \propto v^4$ for a given \gr\ opacity and late--time decay time. A particular supernova 
can only have one $M_{ej}$ and one $E$, so both of these sets of constraints need to 
be at least approximately satisfied. Since the peak relations predict $E \propto M_{ej}^3$ 
and the tail relations predict $E \propto M_{ej}^2$, there is a conflict. Poznanski (2013)
makes an empirical case that the energy scales as the cube of the progenitor mass
for SN~IIP, but no current core--collapse simulations predict $E \propto M_{ej}^2$ or 
$E \propto M_{ej}^3$.

The resolution of this peak/tail conflict may involve consideration of the mean UVOIR opacity 
near the peak, $\kappa$. We note from Equations \ref{masspeak} and \ref{kepeak} that
the observed parameters do not constrain the ejected mass and energy but only
the degenerate combinations $\kappa M_{ej}$ and $\kappa E_{ke}$. By invoking
the constraint that the ejected mass determined from peak parameters be equal to
the ejected mass constrained from tail parameters, the mean opacity can be formally 
expressed in terms of constants ($\beta$ and $C$ depend slightly on density structure),
the reasonably well--known value of $\kappa_{\gamma}$, and of the observed rise time, 
photospheric velocity near peak, and the decay time as
\begin{equation}
\kappa = \frac{5}{3}\beta c C \frac{\kappa_{\gamma} }{v_{ph}}\left(\frac{t_{r}}{T_0}\right)^2.
\label{opacity1}
\end{equation}
For typical parameters this can be expressed as
\begin{equation}
\kappa = 0.01~{\rm cm^2~g^{-1}} \frac{(\kappa_{\gamma}/0.03 {\rm cm^2~g^{-1}})}{v_{ph,9}}\left(\frac{t_{r,10}}{T_{0,100}}\right)^2.
\label{opacity2}
\end{equation}
where $t_{r,10}$ is the rise time in units of 10d and $T_{0,100}$ is the tail decay time in units 
of 100d. We emphasize that this expression for the opacity is basically determined by the 
observed parameters, $v_{ph}$, $t_r$, and $T_0$. 

Equations \ref{masspeak}, \ref{kepeak}, \ref{masstail}, and \ref{ketail} suggest other 
interesting relations. For a given ejected mass and energy, there is a degeneracy
between $\kappa$ and the rise time, $t_r$, such that one can write
\begin{equation}
\frac{\kappa}{t_r^2} = \left(\frac{5}{6}\right)^{1/2}\beta c \left(\frac{C \kappa_{\gamma} E_{ke}}{T_0^2}\right)^{1/4} M_{ej}^{-1}.
\label{opacityME}
\end{equation}
For events with similar rise times and decay times, this implies that $\kappa \propto E_{ke}^{1/4} M_{ej}^{-1}$.

Using Equations \ref{masstail} and \ref{ketail}, one can also write
\begin{equation}
v_{ph} = \left(\frac{10}{3} C\kappa_{\gamma} M_{ej} \right)^{1/2} T_0^{-1},
\label{vM}
\end{equation}
and
\begin{equation}
v_{ph} = \left(\frac{10}{3}\right)^{1/2} \left(C \kappa_{\gamma} E_{ke} \right)^{1/4} T_0^{-1/2},
\label{vE}
\end{equation}
giving the photospheric velocity (mean ejecta velocity) as functions of the ejected mass and
energy. We note that Equations \ref{vM} and \ref{vE} are degenerate in the quantities 
$\kappa_{\gamma} M_{ej}$ and $\kappa_{\gamma} E_{ke}$. We have assumed that 
$\kappa_{\gamma}$ is less dependent on explosion parameters than $\kappa$, but
have kept the $\kappa_{\gamma}$ dependence explicit to remind of that dependence.

{\bf Note added in proof:} {\sl A small, but significant variation in the arguments we present 
here would be to consider the mass given by Equation \ref{masspeak} to not represent 
$M_{\rm ej}$, but the quantity $M_{\rm diff}$, the mass involved in diffusion for which the 
opacity $\kappa \sim 0.1$ \cm2g\ is relevant. From this perspective, the LHS of Equations 
\ref{opacity1} and \ref{opacity2} would become 
$\kappa (M_{\rm diff}/M_{\rm ej})$ and hence
\begin{equation}
\frac{M_{diff}}{M_{ej}} = \frac{5}{3}\beta c C \frac{\kappa_{\gamma} }{\kappa v_{ph}}\left(\frac{t_{r}}{T_0}\right)^2,
\label{mratio1}
\end{equation}
or
\begin{equation}
\frac{M_{diff}}{M_{ej}} = 0.1~{\rm cm^2~g^{-1}} \frac{(\kappa_{\gamma}/0.03 {\rm cm^2~g^{-1}})}
{(\kappa/0.1 {\rm cm^2~g^{-1}}) v_{ph,9}}\left(\frac{t_{r,10}}{T_{0,100}}\right)^2.
\label{mratio2}
\end{equation}
For $\kappa \sim 0.1$ cm$^2$ g$^{-1}$, Equation \ref{mratio2} would then give 
$M_{\rm diff}/M_{\rm ej} \sim 0.1$ for fiducial parameters, strongly suggesting that
only a rather small portion of the ejecta is ionized. With this redefinition of 
the mass from Equation \ref{masspeak}, corresponding changes should be made in 
Equations \ref{T02}, \ref{opacityME}, \ref{vM}, and \ref{vE}.}

The mean opacity derived in this way tends to be $\sim 0.01$~\cm2g\ for events with typical 
peak properties and a late--time decline of $T_0 \sim 160$~d or $\sim 0.016$~\magd\ that 
characterizes the various stripped--envelope events considered in the examples above; this 
is considerably smaller than the estimates normally applied in this context, $\kappa \sim 0.1$~\cm2g. 
We note that the two--component model fit of \citet{vinko04} to SN~2002ap required a mean opacity 
of order 0.01 cm$^{2}$ g$^{-1}$ to fit the observed evolution of the photospheric velocity. If this 
analysis, constrained by the decay properties of the tail, is reasonably correct, then the effective 
opacity that determines the peak properties may be quite low. Many estimates of the ejected 
mass and energy in the literature that assume or compute a higher opacity could be too low by 
as much as a factor of 10.

This analysis shows that the mean opacity that dictates the peak properties of stripped-envelope
supernovae may have been over--estimated in many cases. If so, this would help to resolve 
dilemma (2) outlined above and suggest that results derived from the peak alone may be suspect. 
This analysis has not resolved the other dilemmas, but suggests some new perspectives to bring
to the study of stripped--envelope and perhaps other core--collapse supernovae, suggesting
correlations between the energy and ejected mass required to account for the tail behavior
and how the opacity and expansion velocity might depend on the ejected mass and energy.

\begin{center}

\begin{deluxetable}{lcccccc}
\tabletypesize{\footnotesize}
\tablewidth{0pt}
\tablecaption{Estimates of Ejecta Mass, Energy, and Mean Optical Opacity}
\tablecolumns{7}
\tablehead{
\colhead{Event} & 
\colhead{Type} &  
\colhead{$\left(\frac{\kappa}{0.1~{\rm cm^2 g^{-1}}} \right)M_{ej}$$^1$} & 
\colhead{$\left(\frac{\kappa}{0.1~{\rm cm^2 g^{-1}}} \right)E_{ke}$$^2$} & 
\colhead{$M_{ej}~^3$} & 
\colhead{$E_{ke}~^4$} & 
\colhead{$\kappa~^5$} \\ 
\colhead{SN}  &
\colhead{}  & 
\colhead{$M_{\rm \odot}$} & 
\colhead{$10^{51}$ erg} & 
\colhead{$M_{\rm \odot}$} & 
\colhead{$10^{51}$ erg} & 
\colhead{\cm2g}   
}
\startdata
    1983N    & Ib     & $1.5\pm0.5$              & $0.56\pm0.2$       &   $< [13\pm6]$$^*$       &  $< [4.6\pm2]$$^*$      & $> [0.012\pm0.007]$$^*$ \\
    1983V    & Ic      & $4.4\pm1.3$             & $5.1\pm1.7$          &   $< [34\pm11]$$^*$     &  $< [39\pm14]$$^*$     &  $> [0.013\pm0.006]$$^*$ \\
    1984L    & Ib      & $3.1\pm0.9$              & $1.9\pm0.6$         &   $< [330\pm190]$$^*$ &  $< [200\pm120]$$^*$ &  $> [0.0010\pm0.0006]$$^*$  \\
    1990B    & Ic      & $3.4\pm1$                & $2.5\pm0.8$          &   $12\pm4$            &   5.1                      &  0.047     \\
    1993J     & IIb    & $1.7\pm0.3$              & $0.57\pm0.1$       &   $< 7.7\pm2.4$     &  $< 2.6\pm0.9$      &  $> 0.022\pm0.007$     \\
    1994I     & Ic      & $0.72\pm0.13$          & $0.57\pm0.1$       &   4.2                        &  3.3                       &  0.017     \\
    1996cb  & IIb     & $4.4\pm0.8$              & $2.1\pm0.5$         &   $< 14\pm7$          &  $< 6.8\pm3$        &  $> 0.031\pm0.015$    \\
    1998bw  & Ic     & $2.3\pm0.4$               & $2.6\pm0.6$         &    21                       & 25                         &   0.010    \\  
    1999dn  & Ib     & $1.7\pm0.5$               & $1.0\pm0.3$         &   $< 36\pm14$       &  $< 22\pm9$            & $> 0.0048\pm0.002$   \\
    2002ap  & Ic      & $0.94\pm0.3$             & $1.3\pm0.4$        &   34                        &  47                          &  0.0027   \\
    2003jd   & IIb     & $4.7\pm1.4$              & $6.3\pm2$            &   $< [95\pm80]$$^*$     &  $< [127\pm110]$$^*$  &  $> [0.0049\pm0.004]$$^*$   \\
    2004aw  & Ic      & $15\pm5$                  & $23\pm8$            &    $< [82\pm90]$$^*$      &  $< [125\pm140]$$^*$   &   $> [0.019\pm0.02]$$^*$  \\
    2007Y    & Ib      & $1.6\pm0.3$              & $0.76\pm0.2$      &    $< 7.9\pm1.5$     & $< 3.8\pm0.9$       &  $> 0.020\pm0.004$     \\
    2007gr   & Ic      & $1.1\pm0.2$               & $0.27\pm0.06$    &    $< 7.6\pm2$       &  $< 1.9\pm0.6$      &  $> 0.015\pm0.004$     \\
    2008D    & Ib      & $3.1\pm0.5$              & $1.9\pm0.4$        &    $< 8.6\pm2$        &  $< 5.1\pm1.6$      &   $> 0.036\pm0.01$    \\
    2008ax   & Ib      & $3.0\pm0.5$              & $1.1\pm0.3$        &    $< 15\pm6$        &   $< 5.6\pm2.3$      &   $> 0.020\pm0.008$    \\
    2009bb   & Ic      & $1.1\pm0.3$              & $2.2\pm0.7$        &    $< [7.9\pm7]$$^*$      &  $< [15\pm13]$$^*$      & $> [0.014\pm0.013]$$^*$      \\
    2009jf     & Ib     & $4.9\pm0.8$               & $2.9\pm0.6$        &    $< 74\pm30$       &  $< 44\pm19$        &  $> 0.0065\pm0.003$   \\
    2011bm  & Ic      & $12\pm2$                  & $4.0\pm0.9$        &     $< 40\pm21$      &  $< 13\pm7$           &  $> 0.029\pm0.016$    \\
    2011dh   & IIb     & $2.6\pm0.4$                & $0.74\pm0.16$    &    $< 16\pm5$           &  $< 4.5\pm1.6$      &  $> 0.016 \pm0.006$    \\
\enddata
\vspace{1mm}
$^*$\footnotesize{Entries in square brackets are especially uncertain; see text} 

$^1$\footnotesize{Equation 8}; $^2$\footnotesize{Equation 9}; $^3$\footnotesize{Equation 10}; 
$^4$\footnotesize{Equation 11}; $^5$\footnotesize{Equation 13}
\label{tab:MEkappa}
\end{deluxetable}

\end{center} 

We summarize the implications of these considerations for the supernova sample we
analyze here in Table \ref{tab:MEkappa}. Columns 3 and 4 give the estimates of the 
ejected mass and energy derived from peak properties, scaled to a value of the mean 
opacity of 0.1~\cm2g, from Equations \ref{masspeak} and \ref{kepeak}. Columns 5, 6, 
and 7 give the ejected mass and energy according to Equations \ref{masstail} and \ref{ketail} 
and the opacity according to Equation \ref{opacity1} that would be required to yield the 
latter values of mass and energy. We have propagated the uncertainties in the observed 
parameters $v_{ph}$, $t_r$, and $T_0$ to express the uncertainties in the various derived 
quantities.  As noted in \S \ref{anal}, we formally assume that the uncertainties in these
observed quantities are uncorrelated. Equations \ref{opacityME}, \ref{vM}, and \ref{vE}, 
for instance, suggest that this may not be true. We reserve a deeper study of this issue 
for future work.

As also noted in \S \ref{anal}, our previous determinations of the late--time decay 
parameter, $T_0$, are probably overestimates because we treated the leakage of positrons 
as similar to that of gamma--rays. While positrons may not be completely trapped, there is 
evidence that they are at least partially so in SN~1994I (Clocchiatti et al. 2008). The uncertainty 
in the positron deposition function therefore leads to an intrinsic uncertainty in the quantitative 
value of $T_0$ that comes in squared in Equations \ref{masstail} and \ref{ketail}. Assuming that 
positrons are completely trapped would lead to an underestimate of $T_0$ if they are only 
partially trapped. 

We have accounted for the uncertainty in positron trapping and other uncertainties in the 
following way in Table \ref{tab:MEkappa}. For events for which our uncertainties in the 
observed value of $T_0$ were large (SN~1984L, SN~2003jd, SN~2004aw, SN~2009bb) 
we take the results for ejected mass, energy, and mean opacity based on the slope of the 
tail to be essentially undetermined, but present the values formally derived in square 
brackets. We also regard events without decent R--band or quasi--bolometric light curves 
(SN~1983N, SN~1983V, SN~1984L) to be ill--undetermined in this context and also present 
the associated data in square brackets. For events for which we have determined a late--time 
decay here with an exaggerated treatment of positron leakage, the majority of the events, 
we take the results to be upper limits on ejecta mass and energy and a lower limit on the 
mean opacity. Finally, for the few events for which Clocchiatti et al. (2008) estimated $T_0$ 
assuming complete trapping (SN~1994I, SN~1998bw, SN~2002ap, SN~1990b), we give 
estimates based on those values of $T_0$ (65d, 120d, 142d, and 88d, respectively), recognizing 
that these could be lower limits on the ejecta mass and energy and upper limits on the mean opacity.
Clocchiatti et al. do not give uncertainties for their values of $T_0$ so we have not assigned
formal uncertainties for these four events. From the uncertainties in $v_{ph}$ and $t_r$
alone, there would be uncertainties in the derived quantities of the order of 10 to 20\%. 

The limits we derive for SN~1984L are quite extreme, as follows from the large value of 
$T_0 \sim 660$d we find to fit the flat tail. As the ejecta mass increases, it should become 
more difficult to determine a precise value of $T_0$. The standard deviation we find, 109 d, 
is substantial but still rather small compared to the best--fitting value. At the 3$\sigma$ lower limit 
on $T_0$, the estimates would be less by about a factor of 4, but still large. We may have 
underestimated the uncertainty here. In any case, the propagated uncertainties from all the
observed values make the estimates for SN~1984L from the tail properties indeterminate.
As noted above, the large uncertainties on SN~2003jd, SN~2004aw, and SN~2009bb
mean that the formal results for them are also of little import. For the other events with
upper limits to the ejecta mass and energy in Table \ref{tab:MEkappa}, there is a 
suggestion of considerable spread in derived values. 

Of special interest are the results for the events where the observed values of $T_0$ were
determined in the context of a model that included positron trapping. For SN~1990B,
the results for the ejecta mass and energy were of order a factor of 2 larger using
the tail properties rather than the peak properties with the chosen mean opacity.
For SN~1994I, the discrepancy was even larger, about a factor of 5. The cases
of greatest interest were the broad--lined events, SN~1998bw and SN~2002ap, where
the discrepancy in prediction is an order of magnitude or more.

\section{Conclusions}  
\label{conclusions} 

This analysis has enunciated four dilemmas associated with the late--time light curves
of stripped--envelope supernovae. (1) There is more heterogeneity in the tail
properties than suggested by analyses of the peak alone. (2) The properties derived 
from the peak are often inconsistent with the properties of the tail. (3) Events with 
different spectral types can have rather similar tails. (4) Both typical SN~Ic and 
broad--lined SN~Ic can have similar late--time decay despite the implied differences 
in energy. 

The values for ejecta mass and energy derived from the peak and the parameters that 
determine them frequently predict too steep a late--time decline, especially in 
simple models that assume a constant opacity over the peak of $\sim 0.1$ \cm2g. 
The physics of the late--time light curves is fairly simple, so the discrepancies may lie in 
the physics of the peak, despite the similarity of the immediate post--peak decline determined 
by Drout et al. (2011), Cano (2103), Lyman et al. (2014) and Taddia et al. (2015). Care 
must be taken with values of the ejecta mass and kinetic energy derived with simple 
estimates based only on the properties of the peak.

Our primary goal was not to derive ejecta masses and kinetic energies, rather to caution that 
these quantities can be sensitive to the way the data are selected and employed. Nevertheless, 
we can compare these quantities with related work. We note that Drout et al. (2011) employed 
a measure of the rate of decline to determine estimates of masses and energies, but that this 
time was in some instances shorter than the observed portion of the rise. The rise times we 
present in Table \ref{tab:timescale} are typically a factor of 2 -- 3 larger than those given for 
the same events by Drout et al. Drout et al. also assigned $v_{ph}$ in only two bins, 10,000 \kms\ 
for SN~Ib and normal SN~Ic and 20,000 \kms\ for SN~Ic-BL. This will tend to mute the true dispersion 
of the physical properties of the supernovae that are related to differences in observed velocity. 
Also in contrast, Cano (2013) gives mean values for $v_{ph}$ for SN~Ib of 8027 \kms, for SN~Ic 
of 8470 \kms, and for SN~Ic-BL of 15,114 \kms. Cano does not discriminate between $v_{ph}$ 
and $<v>$ as we have here. For our constant--opacity peak models with $\kappa = 0.1$ \cm2g, 
we tend to assign lower values of $M_{ej}$ and $E_{ke}$ than does Cano (for SN~1983N, SN~1999dn, 
SN~2002ap, SN~2007gr, SN~2008D, SN~2009jf, and SN~2011bm), but agree rather closely in 
several cases (SN~1994I,  SN~2003jd, SN~2007Y, and SN~2009bb). We assign higher values than 
Cano for SN~2004aw. We cannot compare directly to Taddia et al. (2015) because their sample was 
based on SDSS objects not discussed in the general literature. The fits of Taddia et al. were mostly 
for light curves of less than 60 d after maximum. Their fitting procedure also involved a 
construction that employed more parameters than do ours. In a sense, the ratio of our values, 
$T_{0,obs}/T_{0,est}$, is another parameter determined by our process. Lyman et al. (2014) also 
use a measure of the decline rate, rather than the rise, to determine the characteristic light curve 
``width" and employ a two--parameter fitting procedure similar to that of Taddia et al. to fit light 
curve properties up to about 80 d after maximum.

The analysis of the light curves of stripped--envelope supernovae has a long history. \citet{EW88}
invoked clumping as a possible way to address the difficulty of reproducing the peak/tail contrast. 
\citet{CW97} discussed the possibility that the dynamics were in some fashion ``non--homologous" 
with part of the ejecta remaining in a dense core. \citet{Maeda03} constructed a two--component 
model that could fit the data better by dint of having extra parameters. They ascribed the second 
component to a dense inner region, analogous to the ``non--homologous" component of Clocchiatti 
\& Wheeler, to asymmetries in the explosion. Subsequent work has also invoked simple models 
with more parameters and hence more elaborate fits (Cano et al. 2014; Lyman et al. 2014; 
Taddia et al. 2015). The enduring problem has been to get the physics of the peak to agree with 
the physics of the tail. Most suggested solutions have focused on the processes of gamma--ray 
deposition. Here we suggest that the treatment of the opacity may be critical. In truth, the two are 
intimately related and must be treated together and self-consistently. 

A possible factor that could yield a narrower peak for a given ejecta mass and kinetic 
energy and account for the low mean opacity suggested by Equation \ref{opacity2} is 
recombination that would give a time--dependent discontinuity in the opacity, akin 
to that in hydrogen in SN~IIP. \citet{EW88} made the first detailed quantitative models of 
SN~Ib light curves based on helium cores from massive stellar models. They also presented 
one of the few careful exegeses of the nature of recombination fronts and their effect on opacity 
in this context. Whereas many simple models invoke a constant mean opacity over the peak of 
stripped--envelope models, the opacity is very sensitive to temperature, density, and expansion 
effects (H\"oflich, M\"uller \& Khokhlov 1993; Pinto \& Eastman 2000). Interestingly, the models 
of Ensman \& Woosley produced less peak--to--tail contrast than the observations, rather than 
giving too steep a tail decline, the problem we have focused on here. In hindsight, this is because
Ensman \& Woosely employed rather massive helium--star models that retained a substantial 
fraction of the ejecta mass as helium. In their models, all the helium and much of the oxygen 
recombined before maximum, yielding a significant mass that had low opacity and contributed 
little to the optical diffusion time. Most of the opacity over the peak was contributed by the 
relatively low mass of the inner iron and silicon layers. The low effective opacity (Ensman \& 
Woosley mention a typical value of 0.003 \cm2g) allowed them to fit the peak of SN~1983N with 
a rather large ejecta mass, but the whole ejecta mass contributed to the subsequent gamma-ray 
trapping, hence yielding excessively flat late--time light curves. 

Subsequently, models of stripped--envelope light curves have invoked smaller ejecta
masses. Detailed hydrodynamic models that have included complex prescriptions
for the opacity have yielded rather large mean opacities,
$\sim 0.1$ \cm2g. This class of models nearly always found the late--time light curve 
to decline too steeply. Attempts to correct this problem have often addressed adjustments 
to the structure that would affect the gamma--ray trapping. The issue then arises as to whether 
these models that faded too quickly on the tail had, by assumption or computation, too large 
an effective opacity over the peak and hence too small an ejecta mass and too small an 
effective gamma--ray optical depth. 

The UVOIR opacity that governs the shape of the peak is intrinsically complex with dependence 
on density, temperature, and composition and on nonLTE, nonthermal and expansion effects. 
Recombination depends on adiabatic expansion and cooling; subsequent reionization and increase 
in the opacity depends on the deposition of energy from radioactive decay. Recombination can 
lead to a situation where the opacity is a sensitive function of both space and time with the net 
effect being to decrease the effective opacity and thus yield a narrower peak light curve for a 
given ejecta mass. The models of \citet{EW88} amply illustrate that there can be significant 
recombination effects on the mean effective opacity even in spherically--symmetric models. 
In the context of SN~1994I, \citet{KK14} also noted that the recombination of oxygen could lead 
yield a narrower peak (see also Piro \& Morozova (2014). Some numerical models in the 
literature set a floor on the opacity. It is possible that this floor is set too high.

This raises the issue of how much cold, neutral material may exist in the ejecta of various types 
of stripped--envelope supernovae near peak. There must be some excited He~I in both SN~IIb and SN~Ib. 
What is less clear is whether or not there may be a substantial amount of helium that is not 
excited by gamma-ray deposition. \citet{D12} argue that even with an appreciable layer of helium, 
a supernova would not appear as a SN~Ib without outward mixing of \ni. In principle, there could 
be a considerable amount of unexcited helium even in SN~Ic. In each spectral type, and especially 
for SN~Ic, the presence of a significant mass of cold, neutral helium will also affect the spectrum. 
A distinct identifying feature of SN~Ic is the strong absorption corresponding to \OIB. This is one 
of the strongest features in the spectra of SN~Ic, being distinctly stronger than in SN~Ib \citep{M01}. 
Models that include a significant mass of unexcited helium decelerate the oxygen and hence
tend to produce oxygen lines that are too shallow, red, and narrow to correspond to the observed 
feature in typical SN~Ic \citep{D12}.

Asymmetries in the explosions of stripped--envelope supernovae almost surely play a significant role. 
From spectropolarimetry, we know that stripped--envelope supernovae are asymmetric and that 
different elements are ejected at different angles \citep{maund08D}. It is instructive to contemplate 
the image of the SN~IIb Cas A, rent by jets and riddled with irregularities due to nickel bubbles 
\citep{MF13}. This structure is probably a good representation of any realistic stripped--envelope supernova.  

Gamma--ray deposition on the tail is relatively simple, being independent, for instance, of composition 
and relatively insensitive to density irregularities. The physics of the \gr\ deposition around the peak 
depends on the distribution of \ni\ and \co. Any outward ``mixing" of \ni\ would tend to release \grs\ 
more quickly and hence yield a more rapid rise, but then also lead to a more rapid late--time decay, 
which is already a problem with many models. There may be some turbulent ``microscopic" mixing, 
but even Rayleigh--Taylor instabilities tend to form distinct macroscopic plumes in the non--linear 
limit \citep{H10}. The early deposition of gamma--rays will not be azimuthally symmetric and hence 
neither will be the opacity, in contrast to the conditions imposed in any spherically--symmetric model.

The current analysis suggests that the mean opacity around peak is less than frequently assumed or 
computed. The opacity is likely to be dependent on asymmetries that have not yet been adequately 
explored. The small effective opacity we derive suggests that the deposition and heating are irregular 
in real 3D explosions and leave substantial cold, low--opacity, mass near peak that, nevertheless, 
traps \grs\ on the tail.

Consideration of the nature of the opacity near peak might point the way to resolving the second 
dilemma that we have outlined, the self-consistency of the peak and tail properties. This perspective 
may also give some insight into the first dilemma, that there may be more variety in tail behavior 
than is clear at first blush from the peak properties. Perhaps the peak opacity and diffusion time is 
controlled by the fraction of the ejecta comprised of iron-- and silicon--peak elements, and the 
mass of those elements is more uniform than the total ejecta mass. 

It is striking that a substantial subset exists with similar peaks and similar tails comprising events 
of all stripped--envelope spectral types, SN~IIb, SN~Ib, and SN~Ic, and with a range of photospheric 
velocities. Amid the diversity we emphasize, there is a uniformity that demands deeper understanding.
The third dilemma, that stripped--envelope events of different spectral type, and hence different
compositions, ejecta masses and perhaps energies, can have similar late--time light curves
remains a challenge. We have shown that solutions can be found with appropriate choices
of the mean opacity, but that choice implies a scaling with the ejecta mass and energy
(Equation \ref{opacityME}) that is far from obvious. Similar concerns remain for the fourth
dilemma, how both typical and broad--line SN~Ic can have similar late--time light curves. 
The implication is that the energy is somehow correlated with the ejected mass in an 
unexpected way. 

 Our results invoking the variation in the opacity in accord with Equation \ref{opacity2} imply
a rather large range in ejecta masses and energies as given in Table \ref{tab:MEkappa}
and generally larger estimates of ejecta mass and energy than other studies. Taken
at face value, these results would require a substantial reconsideration of the progenitor
evolution. Small ejecta masses are not compatible with the observed class of Wolf--Rayet
stars. Larger ejecta masses might be. Consideration of larger mass progenitors must 
also confront issues of predicted rates of explosions compared to observations and
issues of the sites of the explosions of the various variety of stripped--envelope events
\citep{K14}. Models of the late--time spectra of SN~1985F were more compatible with the 
explosion of a helium core of 8\msun\ than a core of 4\msun\ (Fransson \& Chevalier 1989). 
Some of the issues concerning the light curves of stripped--envelope supernovae might 
be clarified with more investigations of this type to determine the ejecta mass by independent
means.

Even within the simple framework we employ here there are refinements that would 
make a quantitative difference. The ``constants" $\beta$, C, and the ``3/5" that relates 
the mean velocity to the photospheric velocity are functions of the density distribution 
and could be explored for a variety of assumed power--law exponents \citep{vinko04}. 
The nature of positron trapping could be explored more thoroughly. In the future, 
better--sampled light curves from peak to tail would be valuable, and the most accurate 
estimates of the observed UVOIR quasi--bolometric light curve should be employed. 
We do not think these aspects would make a qualitative difference compared to the 
effects of asymmetry, ionization, and recombination we discuss here.
 
Clocchiatti et al. (1997) emphasized the depths of the conundrum presented by events 
with similar late--time tails but different spectral types and different photospheric velocities. 
From Equation (\ref{T01}), two events with the same value of $T_{0,obs}$ must have a (nearly) 
constant value of $M_{ej}^2/E_{ke} = 4 E_{ke}/<v^2>^2$. The similar late--time light curves 
would then require some very unexpected correlation between the kinetic energy and ejecta 
mass and with the the mean, and hence presumably photospheric, velocity. Now, as then, 
the problem posed by the similar light curves and different velocities remains a challenge.

\acknowledgements

We thank Stefano Benetti, Nando Patat, Stefan Taubenberger and Stefano Valenti for sharing their
bolometric light curve data, Dave Arnett and Jozsef Vink{\'o}  for a discussion of Arnett's light curve model,
Maryam Modjaz, Federica Bianco and Mattias Ergon for comments, David Branch for asking a key 
question, and Sean Couch for a valuable perspective. This research is supported in part by 
NSF AST-1109801 and by the Millennium Center for Supernova Science through grant P06-045-F 
funded by ``Programa Iniciativa Cientfica Milenio de MIDEPLAN."

\end{document}